
\input harvmac

\Title{\vbox{\hbox{MRI-PHY/21/95}
\hbox{TIFR-TH/95-52}
\hbox{\tt hep-th/9510233}}}
{Notes on a singular Landau-Ginzburg family}
\bigskip
\centerline{Debashis Ghoshal\foot{E-mail: ghoshal@theory.tifr.res.in.}}
\bigskip\centerline{\it Tata Institute of Fundamental Research}
\centerline{\it Homi Bhabha Road, Bombay 400 005, India}
\smallskip\centerline{\&}
\centerline{\it Mehta Research Institute of Mathematics \&\
Mathematical Physics}
\centerline{\it 10 Kasturba Gandhi Road, Allahabad 211 002, India}
\def\res{\hbox{\rm Res}\,}
\def\sign{\hbox{\rm sign}}
\def\del{\partial}
\def\sssum{{\scriptscriptstyle\sum}}

\vskip .5in
We study some properties of a singular Landau-Ginzburg family
characterized by the multi-variable superpotential
$W=-X^{-1}(Y_1Y_2)^{n-1} + {1\over n} (Y_1Y_2)^n - Y_3Y_4$.
We will argue that (the infra-red limit of) this theory describes the
topological degrees of freedom of the $c=1$ string compactified at $n$
times the self-dual radius. We also briefly comment on the possible
realization of these line singularities as singularities of Calabi-Yau
manifolds.

\Date{\it {October 1995}}

\newsec{Introduction}
Non-critical string theories prove to be instrumental in our
understanding of some of the fundamental issues in highly simplified
situations (see \ref\Mreview{I.\ Klebanov, Lectures at the ICTP Spring
School (1991), hep-th/9108019;\hfill\break
P.\ Ginsparg and G.\ Moore, Lectures at the TASI Summer School (1992),
hep-th/9304011.}\ for a review and reference to the original literature).
Symmetries, behaviour of
the perturbation series to all orders in string coupling, nature of
non-perturbative effects, etc.\ are questions that these `toy' models
have addressed with some success. The simplicity of these theories
derive from the fact that they are conventionally thought of as
strings moving in $d\le 2$ dimensions, where only the topological
(quantum mechanical) degrees of freedom survive. The most interesting
of these is $d=2$: a free scalar field of central charge one on a circle
of radius $R$, coupled to gravity --- the so called $c=1$ string. This
has a propagating massless degree of freedom with momentum quantized
in units of $1/R$.

The topological properties of non-critical string theories are best
described in a formalism that makes them manifest. This is the subject
of topological gravity\ref\WitTG{E.\ Witten, {\it Nucl.\ Phys.} {\bf
B340} (1990) 281.}\ and topological string theory, which for $d<2$
is reviewed in \ref\Dijk{R.\ Dijkgraaf, Lectures at NATO ASI, Carg\` ese
(1991), Eds.\ J.\ Fr\" ohlich {\it et al}, Plenum Press, hep-th/9201003.}.
The same story for
$c=1$ string is more complex. The most symmetric case of $c=1$ string
at the radius $R=1$, (self-dual value under $R\to 1/R$ duality),
variously admits description as double-scaled Penner
model\ref\DiVa{J.\ Distlerand C.\ Vafa, {\it Mod.\ Phys.\ Lett.}
{\bf A6} (1991) 259.},
twisted $SL(2,\hbox{\bf R})_3/U(1)$ coset model\ref\MuVa{S.\ Mukhi
and C.\ Vafa, {\it Nucl.\ Phys.} {\bf 407} (1993) 667.}, topological
Landau-Ginzburg model\ref\GhMu{D.\ Ghoshal and S.\ Mukhi, {\it Nucl.\ Phys.}
{\bf B425} (1994) 173.}\ref\HaOzPl{A.\ Hanany, Y.\ Oz and R.\
Plesser, {\it Nucl.\ Phys.} {\bf B425} (1994) 150.}\ref\GhImMu{D.\
Ghoshal, C.\ Imbimbo and S.\ Mukhi, {\it Nucl.\ Phys.} {\bf B440} (1995)
355.} and its geometrization as a non-compact Calabi-Yau space\ref\GhMuTalk{
D.\ Ghoshal and S.\ Mukhi, in the {\it Proceedings of the International
Colloquium on Modern Quuantum Field Theory}, Bombay (1994), available
at the home page of the Theoretical Physics Group of the Tata Institute.}
and Kontsevich-Penner type matrix model\ref\ImMu{C.\ Imbimbo and S.\ Mukhi,
{\it Nucl.\ Phys.} {\bf B449} (1995) 553.} (see also \ref\DiMoPl{R.\
Dijkgraaf, G.\ Moore and R.\ Plesser, {\it Nucl.\ Phys.} {\bf B394} (1993)
356.}). All these highlight different
aspects of this theory and testify to its rich structure.

Topological string theories also arise from critical strings with
$N=2$ supersymmetry, and are useful in studying a class of
physical operators, the chiral ring, in a simplified scenario which
makes the other operators of the theory BRS-exact\ref\WitTS{E.\ Witten,
in {\it Essays on Mirror Manifold}, Ed.\ S.T.\ Yau, International
Press.}. Many of the interesting physical quantities are tractable in
this framework\ref\BCOV{M.\ Bershadsky, S.\ Cecotti, H.\ Ooguri and
C.\ Vafa, {\it Nucl.\ Phys.} {\bf B405} (1993) 279; {\it Comm.\ Math.\
Phys.} {\bf 165} (1994) 311.}\ref\AGNT{I.\ Antoniadis, E.\ Gava, K. Narain
and T.\ Taylor, {\it Nucl.\ Phys.} {\bf B413} (1994) 162.}\ and using
mirror symmetry\ref\Essays{{\it Essays on Mirror Manifolds}, Ed.\ S.T.\
Yau, International Press.}, leads to some spectacular exact results for
string compactification on Calabi-Yau threefolds.

The topological theories of `toy' non-critical and more physical
critical strings turn out to be intimately related. In \ref\GhVa{D.\
Ghoshal and C.\ Vafa, {\it Nucl.\ Phys.} {\bf B453} (1995) 121,
hep-th/9506122.}\ it was shown that the `Calabi-Yau phase' of
the $c=1$ string at the self-dual radius captures the universal local
geometry (near the nodal point to be) of a generic Calabi-Yau threefold
as it degenerates into a conifold\ref\Conifold{P.\ Candelas, A.M.\ Dale,
C.A.\ L\"utken, and R.\ Schimmrigk, {\it Nucl.\ Phys.}{\bf B298} (1988) 493;
P.S.\ Green and T.\ H\"ubsch, {\it Phys.\ Rev.\ Lett.} {\bf 61} (1988)
1163; {\it Comm.\ Math.\ Phys.}{\bf 119} (1988) 431;
P.\ Candelas, P.S.\ Green, and T.\ H\"ubsch, {\it Phys.\ Rev.\
Lett.}{\bf 62} (1989) 1956; {\it Nucl.\ Phys.}{\bf B330} (1990) 49;
P.\ Candelas and X.C.\ de la Ossa, {\it Nucl.\ Phys.} {\bf B342} (1990)
246.}. A consequence
of this is the fact that the topological free energy of the B-twisted
sigma model\WitTS\ on the Calabi-Yau space has a universal
singular part (as a function of the modulus related to the conifold).
Furthermore, from $c=1$ string theory, this is given by the genus
expansion of its free energy! This in turn determines some of the
couplings of the effective field theory of the type II string and
provides another evidence\ref\AGNTdual{I.\ Antoniadis, E.\ Gava,
K.\ Narain and T.\ Taylor, ICTP Preprint IC-95-177, hep-th/9507115.}\ in
favour of the remarkable non-perturbative duality between type II and
heterotic strings\ref\HIIduality{E.\ Witten, {\it Nucl.\ Phys.}
{\bf B443} (1995) 85 and references therein.}.

In \ref\Strom{A.\ Strominger, {\it Nucl.\ Phys.} {\bf B451} (1995) 96,
hep-th/9504090.}, the physical singularity of the low
energy effective field theory of a string near a conifold was traced
to the infra-red singularity due to the emergence of a new light
(solitonic) black hole degree of freedom. This leads to the tantalizing
possibility of a dynamical non-perturbative process where the topology of the
compactification changes while the underlying string theory remains
smooth\ref\GrMoSt{B.\ Greene, D.\ Morrison and A.\ Strominger, {\it Nucl.\
Phys.} {\bf B451} (1995) 109, hep-th/9504145.}. This
suggests that one should study other singularities of Calabi-Yau
spaces, and as in the case of the simple conifold singularity,
non-critical string theories might help us understand some of the
universal properties of such degenerations.

With this motivation we will propose and study a class of topological
Landau-Ginzburg theories. They are characterized by singular
superpotentials. In Sec.2, we will introduce the models and argue
that they describe the
topological degrees of freedom associated with the $c=1$ string at $n$
times the self-dual radius. In Sec.3, we will identify the `tachyons'
in terms of the Landau-Ginzburg variables and find the selection rule
for correlators, which are then computed in Secs.4 and 5. We find that
while the tree-level correlators agree with the matrix model results,
there is a discrepancy in the genus-1 correlation functions. We will
argue that this is because the Landau-Ginzburg theory describes only
the topological sector which does not include the intermediate tachyons
with fractional momenta. In Sec.6, we discuss their geometry and find
non-compact Calabi-Yau spaces and their singularities. Finally,
in Sec.7, we will briefly comment on their realization as singularities
of compact Calabi-Yau threefolds.

\newsec{The singular family of superpotentials}
Let us first recall the Landau-Ginzburg model for the $c=1$ string at
the self-dual radius\GhMu\HaOzPl\GhImMu. The singular superpotential
$W(X)=-\mu X^{-1}$ is a characteristic feature of this theory and
determines the dynamics of the `tachyon' degrees of freedom
completely. The Landau-Ginzburg theory manifestly has the topological
symmetry algebra with the topological central charge $\hat c=3$.
This value of $\hat c$ is critical in the sense that the free energy
of such theories is naturally well-defined at all genus without any
operator insertion\WitTG. Remarkably $\hat c$ is also 3 for
topological models based on Calabi-Yau spaces, reflecting the fact
that their (complex) dimension is 3. Given the close connection
between Landau-Ginzburg and Calabi-Yau models\ref\LGCY{B.\ Greene, C.\
Vafa and N.\ Warner, {\it Nucl.\ Phys.} (1989) 371;\hfill\break E.\ Witten,
{\it Nucl.\ Phys.} {\bf B403} (1993) 159.}, it is natural
to expect that the $c=1$ string also has a `Calabi-Yau phase'. Indeed
one finds a non-compact analog of Calabi-Yau manifolds\GhMuTalk\
defined by the zero locus of the (augmented) superpotential
\eqn\sdw{W = -\mu x^{-1} + y_1 y_2 - y_3 y_4}
in the non-compact weighted projective space ${\bf P}^4_{-2,1,1,1,1}$.
The additional quadratic terms do not change the value of the central
charge and only make up for the missing coordinates. Nor do they
affect any result obtained for the tachyon correlators using the
superpotential $1/X$.

The non-compact Calabi-Yau space given by the eq.\sdw\ describes the
universal local geometry (near the would be nodal singular point)
of a generic (compact) Calabi-Yau
threefold as it degenerates into a conifold\GhVa. This is apparent
in the coordinate patch $(x\ne 0)$ where we find
\eqn\coni{y_1y_2 - y_3y_4 = \mu}
In the above cosmological constant $\mu$ of the $c=1$ string is the
complex structure modulus that tunes us to the conifold singularity at
$\mu=0$.

The conifold locus \coni\ also has a well known interpretation in
the $c=1$ string theory. The string BRS cohomology of these theories
has an infinite number of non-standard physical operators of zero ghost
number\ref\BRS{B.\ Lian and Zuckerman, {\it Phys.\ Lett.} {\bf B266} (1991)
21;\hfill\break S.\ Mukherji, S.\ Mukhi and A.\ Sen, {\it Phys.\ Lett.}
{\bf B266} (1991) 337;\hfill\break P.\ Bouwknegt, J.\ McCarthy and
K.\ Pilch, {\it Comm.\ Math.\ Phys.} (1992) 541.}. By virtue of being
in the BRS cohomology they also have
dimension zero. The product of such operators is therefore free of
singularity and forms a polynomial ring, called the ground
ring\ref\WitGR{E.\ Witten, {\it Nucl.\ Phys.} {\bf B373} (1992) 187;\hfill
\break E.\ Witten and B.\ Zwiebach, {\it Nucl.\ Phys.} {\bf B377} (1992)
55.}. This ring and the associated variety govern the theory and its
symmetries. Conifold \coni\ is the complexification of this ground
variety\GhVa.

The ground ring and symmetries of $c=1$ string have been studied
as a function of the radius of compactification, and more generally at
all points in the moduli space of the $c=1$ conformal field
theory\ref\GhJaMu{D.\ Ghoshal, D.\ Jatkar and S.\ Mukhi, {\it Nucl\ Phys.}
{\bf B395} (1993) 144.}. At a set of special points --- called the $ADE$
points --- in the CFT moduli space, they are related to the Kleinian
singularities\ref\Klein{F.\ Klein, {\it Lectures on the icosahedron and
solution of equation of the fifth degree}, [English translation (1956)],
Dover.}\ref\Slo{P.\ Slodowy, in {\it Lecture Notes in Mathematics} 1008,
Springer (1983).}. In this paper,
we will only be concerned with the $A_n$-type singularities which
describe the $c=1$ theory compactified on $n$ times the self-dual
radius ($R=n$ in suitable units). For them the locus of the ground
variety bears a close resemblance to eq.\coni. However, we will see
that there are some essential differences between $R=1$ and $R=n>1$.

Recall that the (unperturbed) ground ring describes the singular variety
\eqn\nrgr{{1\over n}(y_1y_2)^n - y_3 y_4 = 0}
We will assume that this is related to a multi-variable superpotential
just as eq.\coni\ is to \sdw. Unlike the theory at the self-dual
radius, the $y$-part is not quadratic and hence contributes to the
central charge. It is easy to work out the U(1) charge assignment for
the Landau-Ginzburg fields corresponding to the coordinates $y$'s such
that the superpotential has charge unity. Let $q_i$ denote the U(1)
charge of the field $Y_i$, then
\eqn\charge{q_1=q_2={1\over 2n}\qquad\qquad q_3=q_4={1\over 2}}
The contribution to the topological central charge of these fields is
therefore ${\hat c}_Y = \sum (1-2q_i) = 2(n-1)/n$. The total topological
central charge of the $c=1$ string is independent of the radius of
compactification\ref\Chat{B.\ Gato-Rivera and A.\ Semikhatov, {\it Phys.\
Lett.} {\bf B288} (1992) 38;\hfill\break M.\ Bershadksy, W.\ Lerche, D.\
Nemeschansky and N.\ Warner, {\it Nucl.\ Phys.} {\bf B401} (1993)
304.}\MuVa\ and hence is $\hat c=3$ for all integer $n$.
The deficit therefore has to be made up by an additional superfield
$X$ with
\eqn\chargex{q_X=-1/n}
that contributes ${\hat c}_X = (n+2)/n$ to the central charge.

This brings us to a singular Landau-Ginzburg family characterized by
the superpotential
\eqn\nradw{W = -\mu X^{-1} (Y_1Y_2)^{n-1} + {1\over n} (Y_1Y_2)^n -
Y_3Y_4}
generalizing the $n=1$ case given in eq.\coni above. This is of course
not the most general possible form as one can add terms of the
form $t_l X^{-l} (Y_1Y_2)^{n-l}$, ($l=2,\cdots,n$), and even terms
where the powers of $Y_1$ and $Y_2$ differ\foot{Terms where $Y_1$
and $Y_2$ have different powers are related to those where they the
same by the $w_\infty$ symmetry in the case of the $c=1$
string\ref\GhLaMu{D.\ Ghoshal, P.\ Lakdawala and S.\ Mukhi, {\it Mod.\
Phys.\ Lett.} {\bf A8} (1993) 3187.}.}.  All of these are
compatible with the U(1) charge assignment, and polynomial in the $Y$
fields and moreover are related closely to the mini-versal deformation
of the singular ground variety \nrgr\GhLaMu. However, in the first
non-trivial case ($n=2$) at least, explicit
computation shows that the extra term $t X^{-2}$ makes no difference
in the computation of the correlators --- the final result are the
same with or without this term. Although we have no general proof for
the above empirical observation, we will, for simplicity, work
with the superpotential \nradw. The general form will be useful in the
later sections when we discuss the singularities and their possible
realization in compact Calabi-Yau threefolds.

\newsec{Landau-Ginzburg operators and selection rule for correlators}
The standard procedure of determining the BRS invariant physical
operators (chiral ring) of a topological Landau-Ginzburg theory
via the Jacobian ideal does not work with a singular superpotential.
For the theory with $W=1/X$, all positive and negative powers of the
superfield $X^{k-1}$, $k\in{\bf Z}$ were proposed to be the physical
states\GhMu.
These were identified to the `tachyons' of momenta $k$ of the $c=1$
string theory. Recall that at the self-dual radius of compactification
all momenta are quantized to be integer valued.

For the theory defined by the superpotential \nradw, let us consider
the following physical operators
\eqn\tachyon{T_k = X^{k-1} (Y_1Y_2)^{n-1},\qquad k\in{\bf Z}}
as a simple generalization of the above idea.
The genus-$g$ correlation function of $N$ such operators $\langle
T_{k_1}\cdots T_{k_N}\rangle$ could be non-zero only if the U(1) charge
conservation
\eqn\conserv{\sum_{i=1}^N (q(T_{k_i})-1) = (g-1)(\hat c - 3)}
law holds. For the critical topological theory we are considering
$\hat c=3$, and the RHS of the above vanishes. Substituting for the
U(1) charge $q$, we find the conservation law
\eqn\momcons{\sum_{i=1}^N k_i = 0}
independent of genus $g$, exactly as in the theory at the
self-dual radius.

This suggests that the operators \tachyon\ are the tachyons of the
$c=1$ string at radius $R=n$. However, the allowed values of momenta
at this radius are in units of $1/n$. The operators $T_k$ in
eq.\tachyon\ cannot therefore account for all the tachyons of the
theory. They describe only a subset of tachyons, namely those with
integer momenta. Interestingly, this subset is special in the sense
that additional discrete states appear precisely at these momenta. The
rest of the tachyons are in the `twisted sector'. They do not have any
discrete state associated with them\GhJaMu.

The collection of integer momentum tachyons and the associated discrete
states are remnants of the higher modes of the string\Mreview,
and form a topological sector of the $c=1$ string
at radius $R=n$. We conjecture that only these topological degrees of
freedom are described by the Landau-Ginzburg theory with the
superpotential \nradw.

The physical states \tachyon\ of course do not exhaust the admissible
physical states of the Landau-Ginzburg theory. In general one can
introduce operators of the form
\eqn\discrete{{\cal O}_{p;r,s} = X^{p-1}Y_1^{r+n-1}Y_2^{s+n-1}}
labelled by three integers $p\in{\bf Z}$, $r,s>-n$. The selection rule
for $N$ such states is
\eqn\seldisc{\sum_{i=1}^N \left(p_i-{r_i+s_i\over 2}\right) = 0}
The operator ${\cal O}$ can therefore be thought of as carrying momentum
$k=p-{r+s\over 2}$. It will be interesting to investigate if these are
related to the discrete states of the $c=1$ string.

\newsec{Residue, contact term and genus-0 correlators}
The simplest correlation function to evaluate is the three-point
function on the sphere. The worldsheet has no modulus and the
three-point function is simply given by the residue formula in the
Landau-Ginzburg theory\ref\DVVLG{R.\ Dijkgraaf, E.\ Verlinde and H.\
Verlinde, {\it Nucl.\ Phys.} {\bf B352} (1991) 59.}\ref\VaLG{C.\ Vafa,
{\it Mod.\ Phys.\ Lett.} {\bf A6} (1991) 337.}:
\eqn\three{
\eqalign{\langle T_{k_1} T_{k_2} T_{k_3} \rangle_{g=0} &\equiv \res_W
\left( T_{k_1} T_{k_2} T_{k_3} \right)\cr &=
\res_W\left[{T_{k_1}T_{k_2}T_{k_3}\over\prod\del W/\del x_j}\right]\cr
&=\res_W\left[{x^{k_1+k_2+k_3+1} y_1y_2\over(xy_1y_2-(n-1))^2 y_3y_4}
\right]\cr
&=\res_W\left[{xy_1y_2\over(xy_1y_2-(n-1))^2 y_3y_4} \right]
\delta_{k_1+k_2+k_3,0}\cr}}
where in the last step we have used the momentun conservation
condition \conserv.

The multivariable residue is discussed in detail in
\ref\GrHa{P.\ Griffiths and J.\ Harris, {\it Principles of Algebraic
Geometry}, Wiley-Interscience (1978).}. Intutively it is the
coefficient of the simple `pole' in the right variables. To this
end, we notice that after imposing the momentum conservation condition
\conserv, the variables $x,y_1$ and $y_2$ always appear in the
combination $xy_1y_2$. Defining new variables $z_1=xy_1y_2$ and
$z_{i+1}=y_i$ for $i=1,\cdots,4$; we find that there are poles at
$z_1=(n-1)$ and $z_2=z_3=z_4=z_5=0$.
Each of these poles is along a (complex)
codimension one hypersurface of ${\bf C}^5$, which can be encircled
by a one (real) dimensional contour. The direct product of these
contours enclose the critical points where all the poles
intersect. With this prescription the residue in the last step of
eq.\three\ is unity. The three-point function of the tachyons is the
momentum conserving delta function giving the expected answer.

Now we come to first non-trivial case, namely the four-point function
on the sphere. The contribution from the bulk of the moduli space is
given by
\eqn\fourbulk{\eqalign{
\langle T_{k_1}T_{k_2}T_{k_3}T_{k_4} \rangle_{g=0}
\big|_{\hbox{\rm bulk}} &= {\del\over\del t_{k_4}}
\res_{W+t_{k_4}T_{k_4}} \left( T_{k_1}T_{k_2}T_{k_3} \right)
\big|_{t_{k_4}=0}\cr &= (1-k_4)\delta_{\sssum k,0} }}
as one would expect from a topological theory of Landau-Ginzburg
matter\VaLG. But this is a (topological) string theory, and involves
integration over the one dimensional moduli space of a sphere with
four punctures. Therefore, in addition to the above, there are
corrections from the boundaries of the moduli space. This
happens when the location of $T_{k_4}$ `collide' with the other
tachyon insertions, which are fixed by the Killing symmetries of the
sphere. The contribution from the boundaries can be neatly encoded by
defining {\it contact terms} between tachyons\ref\LoLG{A.\ Losev, {\it Theor.\
Math.\ Phys.} {\bf 95} (1993) 595.}.

A suitable choice of contact term turns out to be
\eqn\contactdef{C_W(T_{k_1},T_{k_j}) = \sum_{j=1}^5 {\del\over\del x_j}
\left( {T_{k_1}T_{k_2}\over\del W/\del x_j}\right)_{\sign\;
{\displaystyle q_j} }}
The subscript $\{\sign\; q_j\}=\{-++++\}$ is the mnemonic that in the
first term, after differentiating wrt $x$, we only keep terms with
negative powers of $x$, and similarly the $j$th term is retained only
if $y_j$ appears with positive power. The contact term thus defined is
a generalization of that in \GhMu\ and is easily evaluated to be
\eqn\contact{\eqalign{
C_W(T_{k_1},T_{k_2}) = &\Big( (k_1+k_2)T_{k_1+k_2}\theta(-k_1-k_2)\cr
&~~+\;2n{x^{k_1+k_2-1}(y_1y_2)^{n-1}\over(xy_1y_2-(n-1))} -
2{x^{k_1+k_2}(y_1y_2)^n\over(xy_1y_2-(n-1))^2} \Big) }}
The first term in the RHS above is a tachyon and is exactly the same
as in the $n=1$ case for which the second and third term cancel out.
We will see below that only this term contributes to the tachyon
correlators.

Using the contact term \contact, the boundary contribution when
$T_{k_4}$ collides with, say, $T_{k_3}$ is
\eqn\fourbound{\langle T_{k_1}T_{k_2}C_W(T_{k_3},T_{k_4}) \rangle =
(k_3+k_4)\theta(-k_3-k_4)\delta_{\sssum k,0} }
The non-tachyonic terms in \contact\ have higher poles and do not
contribute.
Adding the bulk and boundary contributions, we get the correlation
function of four tachyons on the sphere
\eqn\four{ \langle T_{k_1}T_{k_2}T_{k_3}T_{k_4} \rangle_{g=0} = (1 -
\hbox{\rm max}|k|) \delta_{\sssum k,0} }
This result agrees with the matrix model answer\ref\KlLo{I.\ Klebanov
and D.\ Lowe, {\it Nucl.\ Phys.} {\bf B363} (1991) 543.}\ apart from
an overall normalization factor of $R=n$. Restricting to the kinematical
configuration where $k_j>0$ for $j\ge 4$, all genus-0 correlators
agree with the matrix model result. A general proof of this can be
worked out along the lines of \GhMu. In other kinematical
configurations, the answers agree in all cases that we have checked.

Two comments are in order here. Firstly, although the genus-0
correlators are insensitive to the radius of compactification, the
fact that the Landau-Ginzburg model motivated from the ground ring of
$c=1$ string reproduces them correctly provides a non-trivial check.
Secondly, by taking the momenta $k$ to be valued formally in
$\hbox{\bf Z}/n$, one gets the genus-0 correlators of all tachyons.
However, the fields $T_k$ with fractional $k$ do not, strictly
speaking, belong to the physical states of the topological theory.

\newsec{Correlation functions on the torus}
We now come to the correlation functions on higher genus Riemann
surfaces. Unfortunately these are difficult to compute in string
theory even for the simple topological Landau-Ginzburg models.

The situation is perhaps best understood for topological Landau-Ginzburg
theory with superpotential $W=1/X$\GhImMu. Tachyons with positive
momenta are primaries and those with negative momenta are to be
thought of as gravitational descendants. More precisely, a tachyon
with negative momentum is thought of a linear combination of
different terms each of which is a product of gravitational and
`matter' degrees of freedom:
\eqn\picture{T_{-k} = \sum_{i=0}^k\prod_{j=1}^i(j-k)
\sigma_i\; T_{-k+i}}
For a given Riemann surface, the conservation law uniquely picks up
one of these terms so that the gravitational part $\sigma_i$ accounts
for the dimension of the relevant moduli space. This is the `picture
changing' hypothesis for the tachyons\GhImMu.

In the appropriate `picture' the correlator factorizes into purely
gravitational and matter contributions. The former is given by
topological gravity\WitTG. The matter part is more complicated. But in
each genus, the contribution from the bulk of the moduli space is
given by the handle operator insertion prescription for topological
Landau-Ginzburg matter\VaLG. The various boundary contributions then
come from contact terms between handles. For the $c=1$ string at the
self-dual radius these combine to give the $W_\infty$
constraints\GhImMu.

As an example consider the one-point function of $T_{-k}$ on the
torus. The moduli space is one dimensional, so the correct insertion
of the tachyon is in the `1-picture'
\eqn\onept{\eqalign{
\langle T_{-k}\rangle_{g=1} &=
(1-k)\langle\sigma_1\rangle_{g=1} \langle T_{-k+1}\rangle_{g=1}\cr
&= {(1-k)\over 24}\res_W\left[{\del^2 W/\del x^2\over\del W/\del x}
T_{-k+1}\right]\cr }}
In the above we have used $\langle\sigma_1\rangle_{g=1}=1/24$ in
topological gravity\WitTG\ and that the handle operator is
$\del^2W/\del x^2$\VaLG. All $1\to N$ correlators on the torus, that is
correlators involving one negative and $N$ postive momentum tachyons
can be obtained from \onept\ by starting with the perturbed
superpotential $W+\sum t_k T_k$, and differentiating it $N$ times in the
appropriate couplings $t_k$'s. In this case, it turns out that \onept\
gives the complete answer and that there is no contribution from the
boundary\GhImMu.

We will assume that the above expression (with appropriate
modification to take into account several variables), is true
in the Landau-Ginzburg theory with superpotential \nradw. To evaluate
the correlator, we need the `1-picture' for the tachyon $T_{-k}$.
The proposal we make for this is the following:
\eqn\onepic{T_{-k} \sim (1-k)\sigma_1 {T_{-k+1}\over T_1} =
(1-k)\sigma_1 x^{-k} }
The appearance of the puncture operator $T_1=(y_1y_2)^{n-1}$ in the
above is necessitated by the U(1) charge conservation. An analogous
situation is the closely related formula
\eqn\eguchi{ C_W(\sigma_s\cdot\phi_i,P) \sim \sigma_{s-1}\cdot\phi_i }
of \ref\Eguchi{T.\ Eguchi, H.\ Kanno, Y.\ Yamada and S.-K.\ Yang, {\it
Phys.\ Lett.} {\bf B298} (1993) 73.}\ for the contact term between a
gravitational descendant and the puncture operator in Landau-Ginzburg
matter coupled to gravity. In the theory with $W=1/X$, the puncture
operator is identity and does not appear explicitly.

The proposed form of the tachyon in `1-picture' can be checked by
evaluating the genus-0 four point function of three positive
and one negative momentum tachyons:
$$\langle T_{k_1}T_{k_2}T_{k_3}T_{-k_4}\rangle_{g=0} =
\langle\sigma_1\rangle_{g=1} \res_W(T_{k_1}T_{k_2}T_{k_3}T_{-k_4+1}) =
(1-k_4)\delta_{\sssum k,0} $$
where, $k_i\ge 0$ for all $i=1,\cdots,4$.

Now we will calculate the one-point function of $T_{-k}$ on the torus.
The handle operator is given by the determinant of the matrix of the
second derivatives of the superpotential\VaLG. Using picture changing,
we have
\eqn\torusone{ \langle T_{-k}\rangle_{g=1} = (1-k)
\langle\sigma_1\rangle_{g=1} \res_W\left[{\det ||\del_i\del_j W||
\over\prod\del_i W}{T_{-k+1}\over T_1}\right] }
Substituting the explicit form of the handle operator
\eqn\handle{\det ||\del_i\del_j W || = -\, 2x^{-5}(y_1y_2)^{3n-5}
(xy_1y_2 - (n-1))((2n-1)xy_1y_2-(n-1)(n-2)) }
we find that
\eqn\onefinal{ \langle T_{-k}\rangle_{g=1} = -\; {(2n-1)\over 12}}
This answer unfortunately does not agree with the one obtained from
matrix model (except for $n=1$), which at $R=n$ is
\eqn\matrixone{ \langle T_{-k}\rangle_{g=1}^{MM}
= -\; {1\over 24}(n+{1\over n}) }
The result of the Landau-Ginzburg model is an integral multiple of
that in the theory at the self-dual radius. In the case of $n=2$, it
is easy to check that the result is the same even with the
superpotential $W=tX^{-2}-\mu X^{-1}Y_1Y_2 + {1\over 2} (Y_1Y_2)^2 -
Y_3 Y_4$, which retains all terms related to the deformation of the
ground ring \nrgr, and this is likely to be the case for all $n$.

Likewise we find the torus two-point function. Start with the
perturbed superpotential and differentiate \torusone\ to get
\eqn\torustwo{\langle T_{-k}T_{k} \rangle_{g=1} = {(2n-1)\over 24}
(1-k) (k^2 - k -2)}
Although the functions involved and their individual contribution to the
residue are different from \torusone, the result, once again, is the same
integral multiple of that at the self-dual radius. While the matrix model
result
\eqn\matrixtwo{\langle T_{-k}T_{k} \rangle_{g=1}^{MM} =
{1\over 24}(1-k)\left(n k^2 - n k - \left(n + {1\over
n}\right)\right)}
again disagrees with \torustwo.

The hypothesis that the Landau-Ginzburg theory describes only the
topological degrees of freedom of the $c=1$ string at the radius
$R=n$, could provide a resolution to the descrepancy. If so, it is not
surprising that we do not obtain the expected answer in this
formalism. To match with matrix model, one needs to take into account
the effect of the intermediate tachyons (of fractional momenta) in the
twisted sector. It is interesting that the difference shows up for the
first time in loop computations where all states contribute.

\newsec{Calabi-Yau phase and singularities}
Landau-Ginzburg models are closely associated to Calabi-Yau
spaces\LGCY\ of (complex) dimension ${\rm dim}_{\bf C} = \hat c$ (for
integer values of the central charge). The zero locus of the superpotential
defines the manifold as an hypersurface in some appropriate projective space.
Witten has argued that the two descriptions can be thought of as different
phases of the same theory\LGCY.

The Calabi-Yau phase of the $c=1$ string theory at the self-dual
radius\GhMuTalk\ is particularly interesting. It describes the
universal local geometry of a generic Calabi-Yau threefold as it
degenerates into a simple conifold singularity\GhVa. A conifold
is Calabi-Yau threefold with an isolated singular point of zero
multiplicity\Conifold. Such canonical singularities are generically
present in the (complex structure) moduli space of threefolds and
moreover are only a finite distance away from a typical point there.
It turns out that near the isolated nodal point on the degenerate
manifold, it assumes the universal form \coni\ with $\mu=0$. Small
non-zero value of $\mu$ corresponds to the nearly degenerate
situation. Thus one identifies the cosmological constant $\mu$ of the
$c=1$ string with the complex modulus that tunes to the conifold. The
Calabi-Yau space related to the $c=1$ string at the self-dual radius
is given by the hypersurface defined by the zero of the superpotential
\sdw. This equation is quasi-homogeneous with the weights
\charge\chargex\ (with $n=1$), and is therefore an equation in the
weighted projective space ${\bf P}^4_{-2,1,1,1,1}$. Notice that this as
well as the Calabi-Yau space, which is exactly the conifold, are
non-compact.

Figuratively speaking, the Calabi-Yau phase of $c=1$ string at the
self-dual radius is an approximation to a threefold near a conifold
point as the harmonic oscillator potential $x^2$ is to a generic
potential at a simple critical point. A consequence of this
approximation of local geomtery by the $c=1$ string is the fact that
the universal physics dominated by the formation of the singularity
is described by the non-critical string theory. In particular,
the singular part of the topological free energy of a B-twisted\WitTS\
sigma model based on any degenerating Calabi-Yau manifold is given by
the free energy of the $c=1$ string as a function of $\mu$\GhVa.
This has an exact expression involving the virtual Euler number of the
moduli space of genus-$g$ Riemann surfaces\ref\GrKl{D.\ Gross and I.\
Klebanov, {\it Nucl.\ Phys.} {\bf B344} (1990) 475.}\DiVa\MuVa\GhImMu.
Geometrically this free energy is related to a certain period of the
holomorphic 3-form of the manifold\Conifold. Physically they
determine, for type II strings, some $F$-terms in the effective field
theory\BCOV\AGNT.

Since the Calabi-Yau phase of the $c=1$ string at the self-dual radius
has a useful realization, it is natural to study the same for the
theory at $R=n$ defined by the superpotential \nradw. Repeating
similar argument, it follows that the Calabi-Yau space is defined by
the zero locus of the superpotential\foot{Here we use the general form
of the superpotential.}
\eqn\pertw{W = {1\over n} (y_1y_2)^n - y_3y_4 -\mu x^{-1} (y_1y_2)^{n-1}
+ \sum_{l=2}^n t_l x^{-l}(y_1y_2)^{n-l} }
Due to the weight assigments \charge\ and \chargex\ of the fields, $W=0$
defines a hypersurface in the weighted projective space
${\bf P}^4_{-2,1,1,n,n}$. This ambient projective space is again
non-compact and also has singularities characteristic of weighted
projective spaces. Similary the hypersurface $W=0$ is non-compact.

In the affine patch $(x\ne 0)$, the Calabi-Yau manifold in question is
given by the (polymonial equation)
\eqn\affine{ {1\over n} (y_1y_2)^n - y_3y_4 -\mu (y_1y_2)^{n-1}
+ \sum_{l=2}^n t_l (y_1y_2)^{n-l} = 0 }
in ${\bf C}^4$. For generic values of the parameters $\mu$ and $t_l$ the
hypersurface \affine\ is non-singular. However, it develops
singularity along codimension one subspace of the parameter space
$(\mu,t_l)$. These in turn intersect among themselves along higher
codimension subspaces. The origin of this parameter space defines the
most singular manifold \nrgr.

There is an important difference in the case $n>1$ compared to $n=1$.
The singularity here is not isolated. The hypersurface \nrgr\ is
singular along a (complex) 1-dimensional subspace. The locus of
singularity is given by
\eqn\singularity{(y_1y_2)^{n-1} = 0\qquad y_3=y_4=0}
which is a pair of intersecting lines in ${\bf C}^2$ defined by $y_1$
and $y_2$. This is a two-sided real 2-dimensional cone with circular
base. This singularity is $(n-1)$-fold degenerate.

As an illustration consider the simplest case, $n=2$. The moduli space
of the singularity is two dimensional with coordinates $\mu$ and $t_2$.
The most singular case is for $\mu=t_2=0$, for which the manifold is
singular along $y_1y_2=0$. For the codimension one locus $t_2=\mu^2/2$,
the line singularity persists, but they are along $y_1y_2=\mu$ in the
$y_1y_2$ plane, and is still one-fold degenerate. The real geometry
of this is that of a hyperboloid of revolution. For $t_2=0$, but $\mu\ne 0$,
there is only an isolated singularity at the origin.

The locus of singularity \singularity\ is also familiar from the $c=1$
string. It is merely the {\it complexification} of the singularity of
the ground variety\GhJaMu, which continues to remain singular under
a perturbation by the cosmological operator\GhLaMu. In $c=1$ string,
the singular locus is identified with the fermi level of the matrix model
free fermions\WitGR\GhJaMu. In the latter description, tachyons are thought
of as small deformations (ripples) on the fermi surface\Mreview. In the
topological description, it is only the intermediate tachyons with
fractional momenta that are localized degrees of freedom on the
singular locus\GhJaMu\GhLaMu. Notice that the multiplicity of degeneracy
of the singularity equals the types of intermediate tachyons in the
twisted sector (tachyons with the same fractional part of momentum).
In the limit $R\to\infty$, the singular locus becomes infinitely
degenerate corresponding to the fact that the intermediate tachyons form
a continuum there. The two-dimensional surface of singularity dominates
the dynamics, and should perhaps be identified with the conventional
target space.

The two sides of the fermi sea are connected when we identify
it to the complex singularity. On the other hand, from matrix model, it
has recently been suggested that both sides of the fermi sea should be
taken into account for a consistent string interpretation\ref\wadia{A.\
Dhar, G.\ Mandal and S.\ Wadia, CERN Preprint CERN-TH-95-186,
hep-th/9507121.}. This brings in states that correspond
to the negatively dressed (`Seiberg disallowed'\Mreview) discrete states
of the Liouville theory. Whereas from the ground ring of $c=1$ string,
these precisely give rise to the deformations considered above\GhLaMu.

\newsec{Realization of the singularities in compact Calabi-Yau spaces}
We will now briefly discuss the possibility of realizing the singularities
described above in compact Calabi-Yau spaces. Indeed these singularities
do occur in compact threefolds, and it is rather easy to find such
examples. The first one we will cite is the ubiquitous quintic. Consider
the hypersurface
\eqn\quintic{(y_1y_2)^2y_5 - (y_3y_4 - \mu y_1y_2)y_5^3 + t_2y_5^5 = 0}
in ${\bf P}^4$. This has the $A_2$ type singularity in the coordinate patch
$(y_5\ne 0)$. Similarly the degree 8 hypersurface
\eqn\athree{{1\over3}(y_1y_2)^3y_5 - \left(y_3y_4 -
\mu(y_1y_2)^2\right)y_5^2 + t_2y_1y_2y_5^3 + t_3y_5^4} = 0
in ${\bf P}^4_{1,1,2,2,2}$ has the $A_3$ singularity.
Other examples like ${\bf P}^4_{1,2,3,6,6}[18]$
($A_5$ type) and ${\bf P}^4_{1,1,1,6,9}[18]$ ($A_8$ type) can be found by
searching among known Calabi-Yau spaces.

These are canonical singularities, and are therefore only a finite distance
away from a generic smooth point of the moduli space\ref\WitComments{E.\
Witten, IAS Preprint IASSNS-HEP-95-63, hep-th/9507121.}. The effective
field theory will again reflect the
singularity of the metric on the moduli space, and it will be interesting
to see if these can be resolved in a way analogous to the picture of \Strom.

The topological one-loop free energy that we have found for the $A_n$
singularity using the singular Landau-Ginzburg model is $F_1=-{2n-1\over 12}
\log\mu$. It will be interesting to check if these numbers are related
to the degenerations of compact Calabi-Yau. One should remember that this
may not be full answer, as it does not agree with the matrix model result.
A related point is that the singular locus $y_1y_2=\mu$ could be part of
a curve of non-zero genus and have a non-zero fundamental
group, and it was found that the chiral ring analysis is not sufficient
in such cases\ref\AsMo{P.\ Aspinwall and D.\ Morrison, {\it Phys.\ Lett.}
{\bf B334} (1994) 79.}.

Finally, the Landau-Ginzburg method discussed here can be extended to the
case of $D_n$, $E_6, E_7$ and $E_8$ type singularities by appealing to
the corresponding ground ring of $c=1$ string theory\GhJaMu. We hope that
the $A_n$ models described here and their extensions will be helpful
in understanding different singularities of Calabi-Yau spaces.

\bigskip
{\bf Acknowledgement:} I am grateful to Camillo Imbimbo, Kirti Joshi, Samir
Mathur, Sunil Mukhi, Kapil Paranjape, Ashoke Sen and Cumrun Vafa for
valuable discussion. Part of this work was done at the Tata Institute of
Fundamental Research. I would like to thank the Theoretical Physics Group
there for hospitality.

\bigskip{\bf Note added:} After completing this note, a recent preprint
\ref\BeSaVa{M.\ Bershadsky, V.\ Sadov and C.\ Vafa, Harvard
preprint HUTP-95/A035, hep-th/9510225.} appeared that analyzes the $A_n$
singularities in detail and relates them to the singularities of
$K3\times T^2$.

\listrefs

\end